\documentclass[preprint,tightenlines,twoside,floatfix]{revtex4}

\usepackage{epsfig}
\usepackage{psfrag}
\usepackage{amsmath}
\usepackage{amssymb}
\usepackage{bm}

\def\picdirectory{.}

\begin{document}
\title{Local heat flux and energy loss in a 2D vibrated granular gas}

\author{Olaf Herbst, Peter M\"uller, and Annette Zippelius}

\affiliation{Institut f\"ur Theoretische Physik,
  Georg-August-Universit\"at, D--37077 G\"ottingen, Germany}

\date{\today}

\begin{abstract}    
  We performed event-driven simulations of a two-dimensional granular
  gas between two vibrating walls and directly measured the local heat
  flux and energy dissipation rate in the stationary state.
  Describing the local heat flux as a function of the coordinate $x$
  in the direction perpendicular to the driving walls, we use a
  generalization of Fourier's law, $q(x)=\kappa \nabla T(x) + \mu
  \nabla \rho(x)$, to relate the local heat flux to the local
  gradients of the temperature and density. This {\it ansatz} accounts
  for the fact that density gradients also generate heat flux, not
  only temperature gradients.  The transport coefficients $\kappa$ and
  $\mu$ are assumed to be independent of $x$, and we check the
  validity of this assumption in the simulations.  Both $\kappa$ and
  $\mu$ are determined for different system parameters, in particular,
  for a wide range of coefficients of restitution. We also compare our
  numerical results to existing hydrodynamic theories. Agreement is
  found for $\kappa$ for very small inelasticities only.  Beyond this
  region, $\kappa$ and $\mu$ exhibit a striking non-monotonic
  behavior.
\end{abstract}

\maketitle

Driven granular gases have attracted much attention in recent years
\cite{noije98c,bizon99b,cafiero00,blair03,hayakawa03}.  This is partly
because in these systems, energy loss in inelastic collisions is
eventually balanced by energy input from the driving so that a
stationary state can be attained.  In physically realistic models, the
driving usually acts at the boundaries of the system, e.g.\ in terms
of shearing forces or vibrating walls. Thus, maintaining a stationary
state requires a subtle and well-balanced mechanism to transfer energy
from the system's boundaries to its interior. The local energy (or
heat) flux and the local energy-dissipation rate are therefore at the
heart of every hydrodynamic description
\cite{hayakawa03,jenkins85b,dufty97,sela98,jenkins02} of the
stationary state of driven granular gases.
  
Fourier's law states for elastic systems that the heat current is
proportional to the local temperature gradient, the proportionality
constant being the thermal conductivity $\kappa$ \cite{chapman70}.
For inelastic systems there is an additional contribution to the heat
current from density gradients \cite{jenkins85b}. The corresponding
transport coefficient $\mu$ has no analog in elastic systems.
Theoretical approaches start from the Boltzmann--Enskog equation to
account for these effects.  Jenkins and Richman \cite{jenkins85b} have
used Grad's moment expansion to compute the heat flux for small
inelasticity, whereas Dufty et al.\ \cite{dufty97} have pushed kinetic
models using a {\it stosszahl ansatz} to simplify the collision
operator.  MD simulations have been performed by Soto et al.\ 
\cite{soto99} to study $\mu$ for a granular gas on a vibrated plane in
the dilute limit.

\begin{figure}
  \begin{center}
    \leavevmode \epsfig{file=\picdirectory/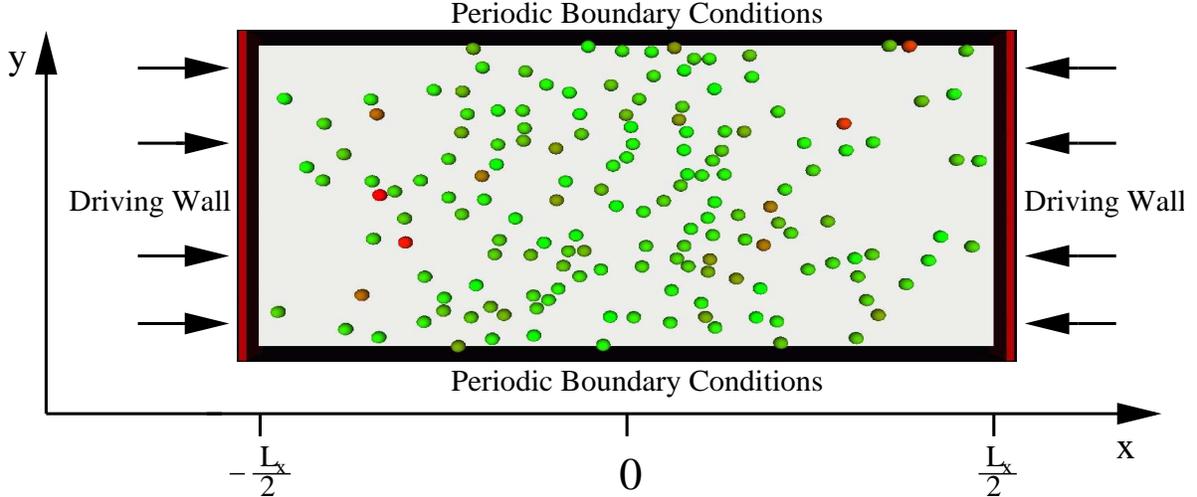, clip=,
      width=0.95\columnwidth, height=0.4\columnwidth}
    \caption{Model of $N$ disks, driven in the $x$-direction with
      periodic boundary conditions in the
      $y$-direction.\vspace*{-0.1cm}
      \label{fig:Schnappschuss}}
  \end{center}
\end{figure}

In this paper we study the local heat flux and local energy-loss
rate, as well as the transport coefficients $\kappa$ and $\mu$ for a
driven granular gas in 2 dimensions.  To do so, we perform
event-driven simulations for the dynamics of $N$ identical inelastic
smooth hard disks of diameter $a$ and mass $m$ which are confined to a
rectangular box with edges of length $L_x$ and $L_y$.  Periodic
boundary conditions are imposed in the $y$-direction, and the gas is
driven through the walls perpendicular to the $x$-direction, see
Fig.~\ref{fig:Schnappschuss} for a typical snapshot.  The left and
right wall vibrate in an idealized saw-tooth manner, characterized by
the driving velocity $v_{\rm drive} >0$: Upon a collision of a
particle with the left/right wall at $x=\mp L_{x}/2$, its
$x$-component of the velocity changes according to $- 2 v_x \pm v_{\rm
  drive}$, see also \cite{herbst04b}.  Inelastic inter-particle
collisions are modeled using a constant coefficient of normal
restitution $\alpha \in\ ]0,1[$ according to $ \hat{\bm{n}} \cdot
\bm{v}'_{12} = -\alpha ~\hat{\bm{n}}\cdot\bm{v}_{12}$.  Here,
$\hat{\bm{n}}$ denotes the unit vector of the particles' relative
center-of-mass positions, and $\bm{v}_{12}$, resp.\ $\bm{v}_{12}'$ are
the pre-, resp.\ post-collisional relative center-of-mass velocities.

A simple estimate in \cite{herbst04b} yields for the spatially
averaged granular temperature $T_{0}=N^{-1} \sum_{i=1}^{N} m
\bm{v}_i^2 /2$ in the stationary state\vspace*{-0.1cm}
\begin{equation}
\label{eq:stat-T} 
  \frac{T_0}{m v^2_{\rm drive}/\varepsilon^{2}}   \approx
  \biggl(\frac{2}{\pi}\biggr)^3 \psi^{-2} \,  
  \left( 1 + \sqrt{1+ (\pi/2)^{2} \varepsilon \psi} \, \right)^2 . 
\end{equation}
Here $\varepsilon := 1-\alpha^{2}$ and $\psi := \sqrt{2} \chi \lambda$
are dimensionless parameters. The latter involves the pair correlation
at contact $\chi$ of the corresponding elastic system and the line
density $\lambda:=N/L_y$.

In the following discussion of heat flux and dissipation we will,
inter alia, be interested in a {\it scaling limit} $\varepsilon
\downarrow 0$ towards an elastic system. In order to prevent the
system from heating up indefinitely when switching off dissipation, we
also need to scale the driving velocity $v_{\rm drive}=:\varepsilon
v_{0}$, where $v_{0}$ is fixed.  Hence, the driving vanishes in the
elastic scaling limit, and the spatially averaged temperature reaches
a finite value, approximately given by $T_{0} \approx (2/\pi)^{3}
(\chi\lambda)^{-2} mv_{0}^{2}$ according to Eq.\ (\ref{eq:stat-T}).

\emph{Locally}, the translational energy changes due to collisions as
well as due to free streaming of the particles in between collisions.
The latter gives rise to a kinetic contribution to the heat current,
$\bm {q}^{\rm kin}$ , whereas the collisions are responsible for the
energy loss $\zeta$ as well as for the collisional contribution to the
heat current $\bm {q}^{\rm int}$. For vanishing macroscopic velocity
the energy balance equation is usually formulated as
\cite{jenkins85b,dufty97}~\vspace*{-0.1cm}
\begin{equation}
  \rho(\boldsymbol{r}, t) \frac{\partial}{\partial t} T(\boldsymbol{r}, t) = -
  \nabla \cdot \boldsymbol{q}(\boldsymbol{r},t) +
  \zeta(\boldsymbol{r},t)
\label{eq:energy.balance.zero.velocity}
\end{equation}
for the hydrodynamic fields of density $\rho$, temperature $T$, total
heat current $\bm {q}=\bm {q}^{\rm kin}+\bm {q}^{\rm int}$ and local
energy dissipation rate $\zeta$.

To compare our simulations to the hydrodynamic theory, we need to
introduce a coarse graining function $\Phi(\bm {r})$ \cite{glasser01},
which is nonzero in a small area centered at $\bm {r}$ only.  We
require of course $\int d\bm {r}\Phi(\bm {r})=1$. In the absence of a
velocity field, the coarse grained kinetic heat current is defined
by~\vspace*{-0.3cm}
\begin{equation}
\label{eq:def.q.kin}
   \boldsymbol{q}^{\rm kin}(\boldsymbol{r},t) =  \sum_{i=1}^{N}
   \frac{m\bm {v}_i^2}{2} \;
   \boldsymbol{v}_i~\Phi{(\boldsymbol{r}-\boldsymbol{r}_i)}~.
\end{equation}
The coarse grained density $\rho(\bm {r})$ and temperature $T(\bm
{r})$ are defined analogously.
 
The change of energy during a binary collision in a small time
interval $\Delta t$ can be decomposed into a source term and a
divergence of a flux.  The coarse grained energy dissipation rate
$\zeta(\boldsymbol{r},t)$ can be calculated analogous to
\cite{glasser01} and is given by
\begin{align}
  \zeta(\boldsymbol{r},t) := \frac{1}{2\Delta t} \;
  \sum_{i,j}{}\raisebox{.7ex}{$'$} (\Delta E_{i|j}+\Delta
  E_{j|i})~\Phi{(\boldsymbol{r}-\boldsymbol{r}_i)}
\label{eq:def.zeta}
\end{align}
in terms of the change of energy $\Delta E_{i|j}$ of particle $i$ due
to a collision with particle $j$ in the time interval $[t, t+\Delta
t]$. The prime at the summation sign restricts $i$ and $j$ to those
particles colliding in $\Delta t$.  The energy dissipation rate
trivially vanishes in the elastic case, when $\Delta E_{i|j}=-\Delta
E_{j|i}$.  Similarly, the collisional contribution to the heat current
is given by
\begin{align}\nonumber
  {\boldsymbol{q}}^{\rm int}(\boldsymbol{r},t) := \frac{1}{4\Delta t}
  \sum_{i,j}{}\raisebox{.7ex}{$'$}
  &(\Delta E_{i|j}-\Delta E_{j|i}) ~\boldsymbol{r}_{ij} \\
  & \times
  \int_{0}^{1}\Phi{(\boldsymbol{r}-\boldsymbol{r}_i+s\boldsymbol{r}_{ij})}~ds
  \,,
\label{eq:def.of.q.int}
\end{align}
where $\bm {r}_{ij}=\bm {r}_i- \bm {r}_j$.  In our simulations we have
never found any significant variations of the long time averages of
the hydrodynamic fields in the $y$-direction parallel to the driving
walls.  Hence we coarse grain our system by subdividing the box into
stripes of width $\Delta x$.

In the following we report numerical results on the stationary state
only. The reader who is interested in more details of the simulations,
such as initialization or relaxation towards the stationary state, is
referred to \cite{herbst04b}.  For dimensional reasons, the driving
velocity $v_0$ sets the time and energy scale and is chosen to be $1$.
Likewise, the particle mass and diameter set the mass and length
scales and are set to $1$, too.

\begin{figure}
  \begin{center}
    \leavevmode
    \epsfig{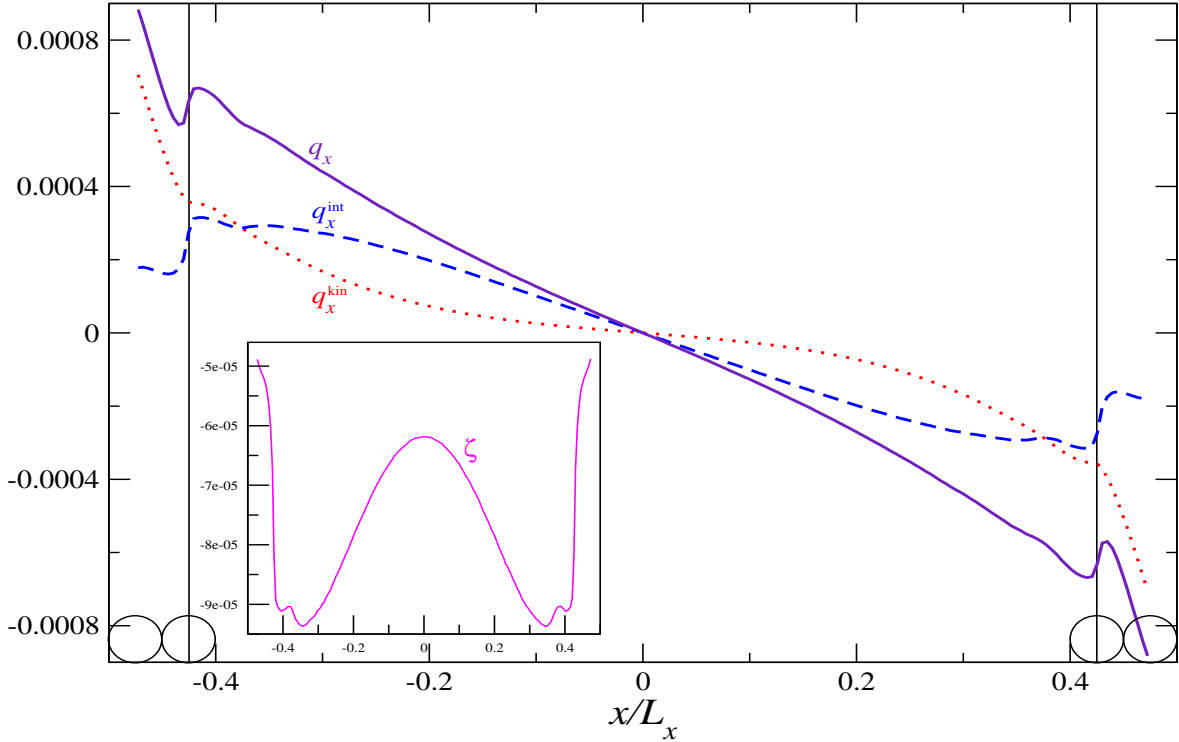}
    \caption{The heat flux in the $x$-direction $q_x$ for a system 
      of $N=256$ particles in a box of size $L_x=20$ and $L_y=25$,
      corresponding to a total area fraction of $\phi_0=0.4$.  The
      dotted line shows the kinetic part $q_x^{\rm kin}$, the dashed
      line is the collisional part $q_x^{\rm int}$, and the full line
      represents the total heat flux $q_x$. The inset displays the
      local energy-loss rate $\zeta$.\vspace*{-0.1cm}
      \label{fig:heatflux}}
  \end{center}
\end{figure}

We first present data for the heat-flux profile in Fig.\ 
\ref{fig:heatflux} for a system of global area fraction $\phi_0 :=
N\pi/(4 L_{x}L_{y}) =0.4$ and coefficient of normal restitution
$\alpha=0.9$.  More precisely, the figure displays the $x$-component
of the heat flux and its kinetic and collisional part as a function of
the normalized position $x/L_{x}$. The inset shows the local
energy-loss rate and will be discussed below.  The heat flux is
antisymmetric about the middle of the system as expected. We clearly
see that its collisional part (\ref{eq:def.of.q.int}), which is
represented by the dashed line cannot be neglected. It is of the same
order as the kinetic contribution---depicted by the dotted line and
corresponding to Eq.  (\ref{eq:def.q.kin}).  For lower density systems
the collisional contribution becomes less important but still it is
not negligible for global area fractions even as low as $\phi_0=0.1$.

\begin{figure}
  \begin{center}
    \leavevmode
      \epsfig{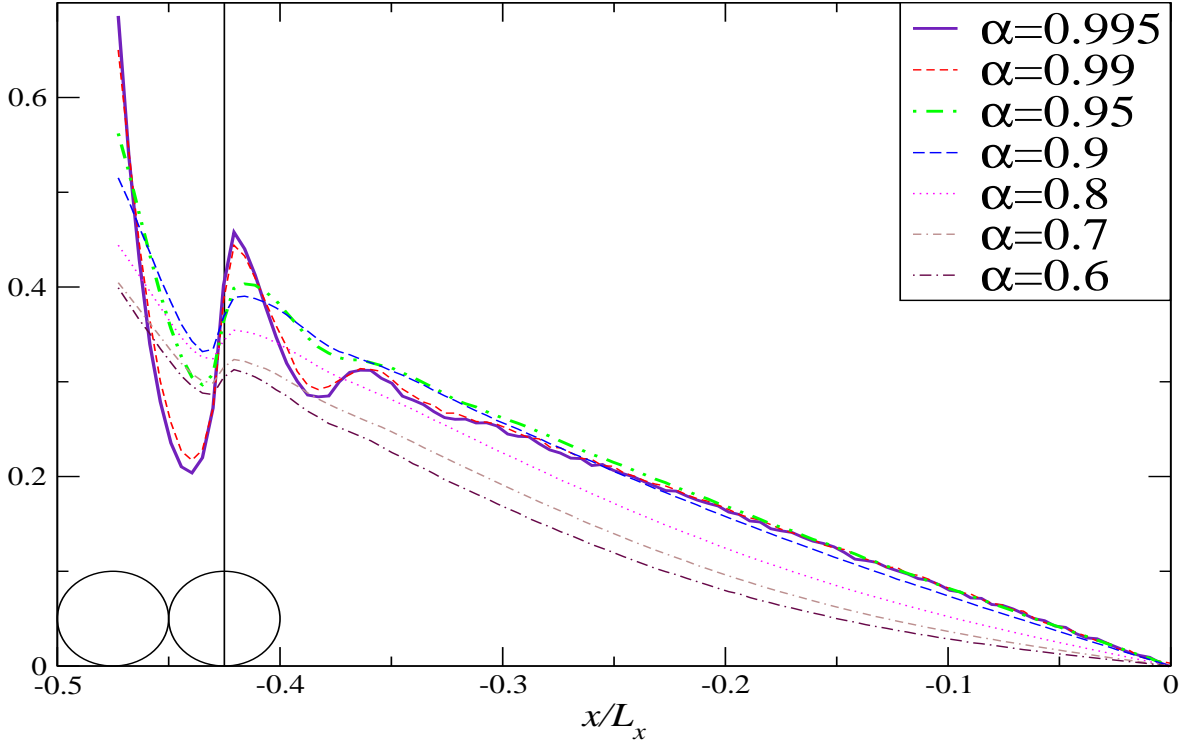}
    \caption{Rescaled heat flux
      $(1-\alpha^2)^{-1}(1+\alpha)^{-2}q_x$ for a wide range of
      coefficients of restitution $0.6 \leq \alpha \leq 0.995$ for
      otherwise fixed systems ($L_x=20$, $N/L_y=10.24$
      [$\phi_0=0.4$]).\vspace*{-0.1cm}
            \label{fig:heatfluxscaling1}}
  \end{center}
\end{figure}

Now we turn to the dependence of the heat flux on the coefficient of
restitution $\alpha$ in the elastic scaling limit. We recall that the
driving strength has been adjusted such that the granular gas attains
a finite temperature as $\varepsilon \to 0$. The heat flux is
proportional to the temperature per time and hence expected to scale
like $T_{0} v_{\rm drive}\sim\varepsilon$, cf.\ (\ref{eq:stat-T}). This
argument is checked by plotting $q_x/\varepsilon$ for different
coefficients of restitution. The collapse of different data sets is
excellent for almost elastic systems ($\varepsilon \ll 1$).  With a
slight modification of the scaling according to $q_x/[\varepsilon
(1+\alpha)^2]$, the data collapse also works approximately for the
full range $0.6 < \alpha < 0.995$ and is shown in Fig.\ 
\ref{fig:heatfluxscaling1}.

\begin{figure}
  \begin{center}
    \leavevmode
    \epsfig{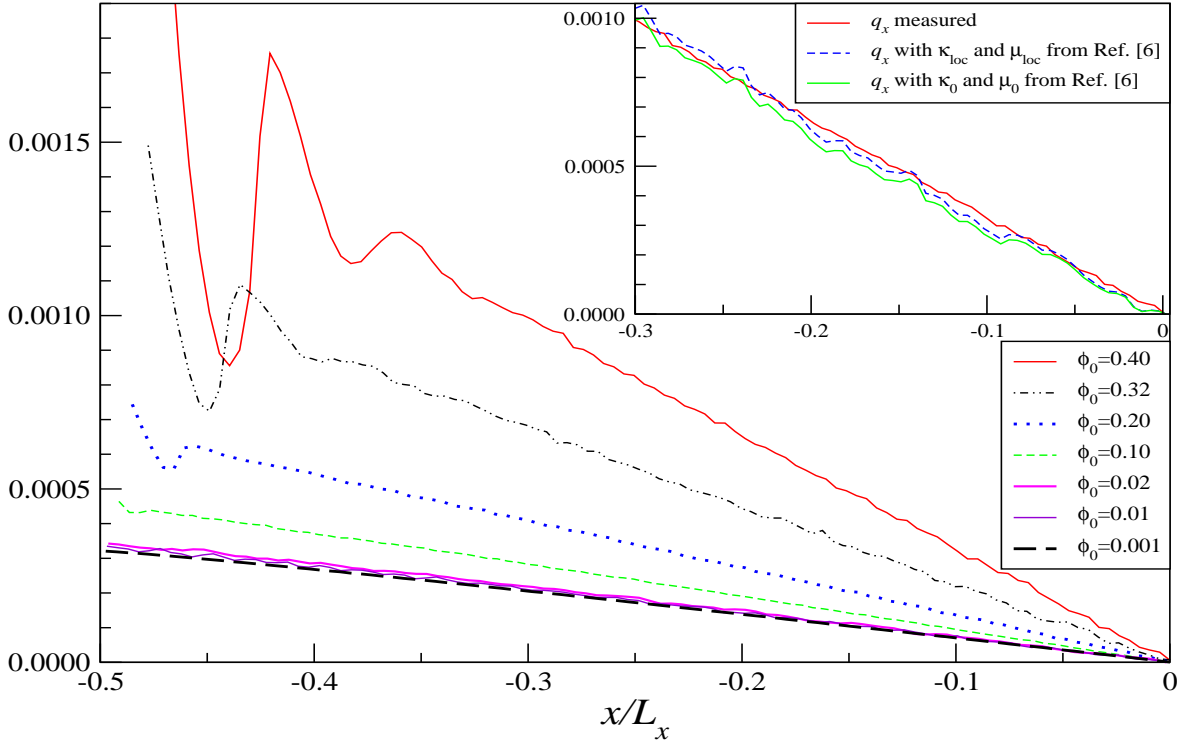}
    \caption{Rescaled heat flux
      $L_x q_x$ for a wide range of box edges (area fraction $0.001
      \leq \phi_0 \leq 0.4$ at fixed line density $N/L_y=10.24$ and
      fixed coefficient of restitution $\alpha=0.99$.  This graph
      looks almost the same for all $\alpha \ge 0.9$.\vspace*{-0.1cm}
            \label{fig:heatfluxscaling2}}
  \end{center}
\end{figure}

In Fig.\ \ref{fig:heatfluxscaling2} we show the heat flux for a fixed
coefficient of restitution $\alpha=0.99$ and various system sizes
$L_x$.  It has been shown \cite{herbst04b} that the scale for
temperature inhomogeneities is set by $L_x$, so that we expect the
heat flux to scale like $L_x^{-1}$, if plotted versus $x/L_x$.  In
Fig.\ \ref{fig:heatfluxscaling2} we plot $q_x L_x$ for various system
sizes $L_x$ and get a decent data collapse for dilute systems.  For
higher densities this scaling captures at least the correct order of
magnitude.

It is a characteristic feature of hydrodynamics of inelastically
colliding particles that a heat current can be generated not only by a
nonuniform temperature but also by density inhomogeneities.  If the
spatial variations of temperature or density are restricted to long
wavelengths, one would expect a gradient expansion to hold. The
simplest constitutive equation for the heat flux is thus a
straightforward generalization of Fourier's law, as discussed in the
literature, cf.~\cite{chapman70}. Chapman--Enskog expansions of both
the Boltzmann and the Boltzmann--Enskog equations predict that the
heat flux of an inelastic system is given by\vspace*{-0.2cm}
\begin{equation}
\label{eq:heatflux}
{q}_{x}(x) = - \kappa \;\frac{d}{dx} T(x) 
+ \mu \; \frac{d}{dx} \rho(x)
\end{equation}
where $\kappa$ is the heat conductivity and $\mu$ is a new transport
coefficient that has no analog for elastic systems
\cite{jenkins85b,dufty97}.

\begin{figure}
  \begin{center}
    \leavevmode
    \epsfig{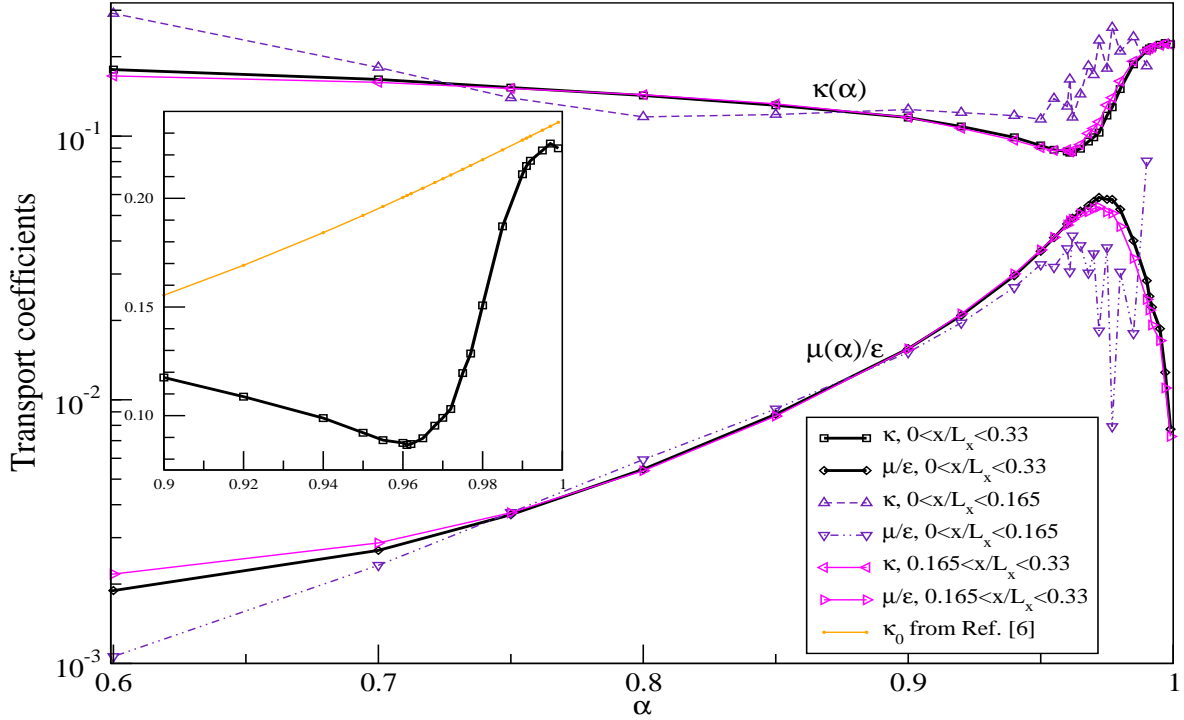}
    \caption{Transport coefficients $\kappa$ and
      $\mu$ as functions of the coefficient of restitution $\alpha$
      for systems of size $L_x=20$ and line density $N/L_y=10.24$
      ($\phi_0=0.4$)\vspace*{-0.1cm}.
            \label{fig:transport}}
  \end{center}
\end{figure}

To estimate the transport coefficients from our data, we assume that
both, $\kappa$ and $\mu$, do not depend on 
the position $x$, i.e.\ we assume them to be constant throughout the
box. It is then straightforward to extract them from a fit of our data
to the above expression (\ref{eq:heatflux}). In Fig.\ 
\ref{fig:transport} we show both transport coefficients $\kappa$ and
$\mu/\varepsilon$.  Both are non-monotonic in $\alpha$.  In the elastic
limit we find that $\kappa$ tends to a non-zero constant, while $\mu
\propto \varepsilon$. The fit of the heat flux as computed from Eq.\ 
(\ref{eq:heatflux}) with constant $\kappa$ and $\mu$ is very good for
all investigated $\alpha$, e.g. for $\alpha=0.9$ it cannot be
distinguished from the data shown in Fig.\ \ref{fig:heatflux}.

Both transport coefficients have been computed within kinetic theory
for small inelasticities by Jenkins and Richman \cite{jenkins85b}, who
find for the thermal conductivity\vspace*{-0.1cm}
\begin{equation} 
\label{eq:Jenkins}
 \kappa=\left[\frac{2 + 3\phi \chi {r}^2(4{r}-3)}{{r}
  \chi(17-15{r})} (2+3\phi \chi {r} )
 +\frac{8\phi^2 \chi {r}}{\pi}\right]\sqrt{\frac{T}{\pi}}\,,
\end{equation}
where $r:=(1+\alpha)/2$.  We use the Henderson approximation
\cite{henderson75} for the pair correlation at contact $\chi$ which is
a function of the area fraction only.  Inserting the local temperature
$T(x)$ and local area fraction $\phi(x)$ from our simulations into
(\ref{eq:Jenkins}), we obtain a spatially dependent transport
coefficient $\kappa_{\rm loc}(x)$ and, from a corresponding equation
in \cite{jenkins85b}, $\mu_{\rm loc}(x)$.  The resulting heat flux
from (\ref{eq:heatflux}) is shown in the inset of Fig.\ 
\ref{fig:heatfluxscaling2} and found to agree very well with the
simulations for $\alpha \geq 0.99$. The agreement is reasonable even
up to $\alpha \sim 0.96$. For larger inelasticities the theoretical
curves capture the correct order of magnitude, but overestimate the
curvature of $q_x(x)$.

Alternatively we can use the global or mean temperature $T_{0}$ and
area fraction $\phi_{0}$ to evaluate Eq.\ (\ref{eq:Jenkins}) and the
respective equation from \cite{jenkins85b} for $\mu$. In Fig.\ 
\ref{fig:transport} we compare this $\kappa_{0}$ to the thermal
conductivity $\kappa$ we obtained from fitting our simulations to Eq.\ 
(\ref{eq:heatflux}). The agreement is good as long as $\alpha \geq
0.99$. As to $\mu_{0}$ (not shown), the deviations to $\mu$ are quite
strong.  Note, however, that the theory of Ref. \cite{jenkins85b} as
well as the simulations of Ref.  \cite{soto99} apply to the low
density limit and not to $\phi_0=0.4$.  Furthermore, the difficulties
in our fitting procedure increase considerably as $\alpha \to 1$
because the temperature gradient becomes proportional to the density
gradient.  Consequently, it is not possible for $\alpha \to 1$ to
determine two parameters from the fit unambiguously.

The difference between $\kappa_{\rm loc}(x)$ and $\kappa_0$ is only a
few percent for $\alpha \geq 0.99$.  The difference in heat flux is
even less, because the strongest inhomogeneities in the transport
coefficients occur in the middle of the sample where the gradients of
temperature and density vanish. The heat flux, computed using the
constant transport coefficients $\kappa_{0}$ and $\mu_{0}$, is also
shown in the inset of Fig.\ \ref{fig:heatfluxscaling2} for comparison.
The fluctuations of the transport coefficients with $x$ increase with
increasing inelasticity, e.g. for $\alpha=0.9$ we find $\kappa_{\rm
  loc}(0)/\kappa \approx 1.5$.

To estimate the degree of inhomogeneity we have divided the box into
an inner and an outer part and fitted the heat current using data from
either half of the box only.  The scattering of the data from the
inner and outer part is shown in Fig.\ \ref{fig:transport} and
provides a rough measure for the effects of inhomogeneous transport.
The results for the outer half are almost identical to the ones for
the full system.  This is also true for the inner part except for the
weakly inelastic systems for which the absolute value of the heat flux
in the middle of the sample is so small that statistical fluctuations
dominate.

The local energy loss, as defined in Eq.\ (\ref{eq:def.zeta}), is
shown in the inset of Fig.\ \ref{fig:heatflux}.  Again, for $\alpha >
0.99$ the agreement with the predicted
$\zeta=-16\varepsilon\phi^2\chi(T/\pi)^{3/2}$ from \cite{jenkins85b}
is very good (not shown). For very dilute quasi-elastic systems
($\phi_0 \lesssim 0.1$, $\alpha \gtrsim 0.95$) the dissipation is
highest in the middle of the sample.  For denser and/or more inelastic
systems we find the absolute value of $\zeta$ (the dissipation) to be
greatest in an intermediate region $x \sim \pm 0.3\, L_x$.  Even
though the density is largest in the middle of the sample, energy
dissipation is not very pronounced there because the mean kinetic
energy is already comparatively small.  Integrated over the whole box,
the local energy loss has to fulfill the conservation law $q_x(-L_x/2)
- q_x(+L_x/2) = \int_{-L_x/2}^{L_x/2}\,\zeta(x) \, dx$.  This, of
course, is confirmed in the simulations, but requires careful
measurements of the heat flux at the boundaries.

In this paper we have discussed the heat current in a granular gas
driven by vibrating walls. We have measured the collisional
contribution to the heat current and have shown that it cannot be
neglected, not even in low density systems.  We have extracted
transport coefficients, the thermal conductivity $\kappa$, as well as
the transport coefficient $\mu$, which accounts for a heat flux due to
density inhomogeneities.  Both transport coefficients have been
determined for a wide range of inelasticities $0.6 \leq \alpha \leq
0.995$ and area fractions $0.01 \leq \phi_0 \leq 0.4$.  We have
studied in detail the elastic limit which is reached by scaling the
driving velocity with $\varepsilon$, so that the temperature tends to a
finite value as $\alpha \to 1$. Comparing our data to theoretical work
in two dimensions \cite{jenkins85b} we found good agreement for
$\alpha \geq 0.99$.  For stronger inelasticities non-monotonic
behavior is observed: The thermal conductivity $\kappa$ first
decreases with increasing $\varepsilon$, goes through a minimum around
$\alpha\approx 0.96$ and then increases again. The transport
coefficient $\mu/\varepsilon$ first increases with $\varepsilon$, goes
through a maximum at $\alpha\approx 0.96$ and then decreases again.
Furthermore a rough estimate of the fluctuations of the transport
coefficients with spatial position in the box has been given. These
effects are expected to be strong for moderately inelastic systems
which are characterized by nonuniform density and temperature and need
further investigation.

We thank Hans Vollmayr for pointing out to us the importance of the
elastic limit and Isaac Goldhirsch for suggesting to measure the heat
flux. We acknowledge financial support by the DFG through SFB~602, as
well as Grant Nos.\ Zi~209/6--1 and Mu~1056/2--1.

\bibliographystyle{apsrev}

\end{document}